\hfuzz 2pt
\font\titlefont=cmbx10 scaled\magstep1

\font\bigmath=cmmi10 scaled\magstep2
\magnification=\magstep1

\null
\vskip 2cm
\centerline{
{\titlefont DISSIPATIVE CONTRIBUTIONS TO}{\hskip 3pt}
{\bigmath \char'042}\hskip 1pt\raise.5ex\hbox{$^\prime$}\hskip -2pt
{\bigmath \char'075\char'042}
}
\vskip 2.5cm
\centerline{\bf F. Benatti}
\smallskip
\centerline{Dipartimento di Fisica Teorica, Universit\`a di Trieste}
\centerline{Strada Costiera 11, 34014 Trieste, Italy}
\centerline{and}
\centerline{Istituto Nazionale di Fisica Nucleare, Sezione di 
Trieste}
\vskip 1cm
\centerline{\bf R. Floreanini}
\smallskip
\centerline{Istituto Nazionale di Fisica Nucleare, Sezione di 
Trieste}
\centerline{Dipartimento di Fisica Teorica, Universit\`a di Trieste}
\centerline{Strada Costiera 11, 34014 Trieste, Italy}
\vskip 2cm
\centerline{\bf Abstract}
\smallskip
\midinsert
\narrower\narrower\noindent
The time evolution and decay of the neutral kaon system can be
described using quantum dynamical semigroups. Non-standard terms appear in
the expression of relevant observables; they can be parametrized
in terms of six, new phenomenological constants. We discuss how the presence
of these constants affects the determination of the parameter
$\varepsilon'/\varepsilon$.
\endinsert
\bigskip
\vfil\eject

The neutral kaon system can be viewed as a specific example
of an open quantum system.[1-3] Its propagation and decay are
described by time-evolutions that generalize the standard approach, 
keeping into account possible new effects due to dissipation and loss of
coherence;[4-6] these extended evolutions form a so-called
quantum dynamical semigroup.

The physical motivation behind this description comes from the
observation that the quantum fluctuations of the gravitational
field at Planck's scale should produce loss of quantum coherence.[7]
The origin of these effects can be ultimately traced to the
dynamics of strings; indeed, one can show that
a generalized, dissipative dynamics for
the neutral kaon system can be the result of its weak interaction
with a gas of D0-branes, effectively described by a heat-bath of
quanta obeying infinite statistics.[8]

Nevertheless, the general structure of the extended 
neutral kaon dynamics based on quantum dynamical semigroups 
is essentially independent from the actual
``microscopic'' mechanism that drives the dissipative effects.
Indeed, the form of the evolution equation is fixed by 
basic physical requirements, like forward in time composition
(semigroup property), probability conservation, entropy increase
and complete positivity. In particular, this last property is
crucial for the consistency of the generalized time evolution,
since it guarantees, also for correlated systems, the positivity
of the probabilities (for a complete discussion, see [9]). 

A rough dimensional estimate reveals that the new, dissipative
effects are expected to be very small, at most of order
$m_K^2/M_P\sim 10^{-19}\ {\rm GeV}$, where $m_K$ is the kaon
mass, while $M_P$ is Planck's mass. Nevertheless, a detailed study 
of the $K^0$-$\overline{K^0}$ system using quantum dynamical semigroups
shows that the dissipative
contributions to the extended dynamics should be in the reach
of the next generation of dedicated kaon experiments.[5, 6]
In view of these results, in the following we shall analyze in
detail how the non-standard dissipative contributions could
affect the determination of the parameter
$\varepsilon'/\varepsilon$.

\bigskip

As usual, we shall model the time-evolution and decay of
the $K^0$-$\overline{K^0}$ system by means of a two-dimensional
Hilbert space.[10] Since we are discussing phenomena leading to
possible loss of quantum coherence, kaon states will be
described by a density matrix $\rho$.
This is a positive hermitian operator, {\it i.e.} with positive eigenvalues,
and constant trace (for unitary evolutions).
In the basis $|K_1\rangle$ and $|K_2\rangle$ of definite $CP$, one thus has:
$$
\rho=\left(\matrix{
\rho_1&\rho_3\cr
\rho_4&\rho_2}\right)\ ,\qquad \rho_4\equiv\rho_3^*\ .
\eqno(1)
$$
The evolution in time of this matrix can be described in general by an equation
of the following form:
$$
{\partial\rho(t)\over\partial t}=-iH_{\rm eff}\, \rho(t)
+i\rho(t)\, H_{\rm eff}^\dagger 
+L[\rho]\ .\eqno(2)
$$
The effective hamiltonian $H_{\rm eff}$ (the Weisskopf-Wigner hamiltonian)
includes a nonhermitian part, that characterizes 
the natural width of the states: $H_{\rm eff}=M-{i\over 2}{\mit\Gamma}$.
The entries of these matrices can be expressed in terms of the 
complex parameters $\epsilon_S$, $\epsilon_L$, appearing in the
eigenstates of $H_{\rm eff}$, 
$$
|K_S\rangle={1\over(1+|\epsilon_S|^2)^{1/2}}
\left(\matrix{1\cr\epsilon_S}\right)\ ,\qquad
|K_L\rangle={1\over(1+|\epsilon_L|^2)^{1/2}}
\left(\matrix{\epsilon_L\cr 1}\right)\ ,
\eqno(3)
$$
and the four real parameters, $m_S$, $\gamma_S$ and $m_L$, 
$\gamma_L$, the masses and widths of the states in (3), 
characterizing the eigenvalues of $H_{\rm eff}$: 
$\lambda_S=m_S-{i\over 2}\gamma_S$, $\lambda_L=m_L-{i\over 2}\gamma_L$.
It proves convenient to use also the following positive combinations:
$\Delta\Gamma=\gamma_S-\gamma_L$, $\Delta m=m_L-m_S$,
as well as the complex quantities:
$\Gamma_\pm=\Gamma\pm i \Delta m$ and
$\Delta\Gamma_\pm=\Delta\Gamma\pm 2i\Delta m$,
with $\Gamma=(\gamma_S+\gamma_L)/2$.

The additional piece $L[\rho]$ generates a trace-preserving, positive,
dissipative map, {\it i.e.} with increasing entropy
$S[\rho(t)]=-{\rm Tr}[\rho(t)\, \log\rho(t)]$,
whose form is uniquely fixed by the requirement of complete positivity.
To write it down explicitly, one expands 
$\rho$ in terms of Pauli matrices $\sigma_1$, $\sigma_2$, 
$\sigma_3$ and the identity $\sigma_0$: 
$\rho=\rho_\mu\, \sigma_\mu$, $\mu=0,1,2,3$.
In this way, the map $L[\rho]$ can be represented by
a symmetric \hbox{$4\times 4$} matrix $\big[L_{\mu\nu}\big]$, 
acting on the column vector with components 
$(\rho_0,\rho_1,\rho_2,\rho_3)$.
It can be parametrized by the six real 
constants $a$, $b$, $c$, $\alpha$, $\beta$ and $\gamma$:[4]
$$
\big[L_{\mu\nu}\big]=-2\left(\matrix{0&0&0&0\cr
                                     0&a&b&c\cr
                                     0&b&\alpha&\beta\cr
                                     0&c&\beta&\gamma\cr}\right)\ ,
\eqno(4)								 
$$
with $a$, $\alpha$ and $\gamma$ non-negative.
These parameters are not all independent; to assure the complete positivity
of the time evolution generated by (2), they have to satisfy
certain inequalities (for details, see [4, 6]).

Physical properties of the neutral kaon system can be computed 
from the density
matrix $\rho(t)$, solution of (2), by taking its trace with
suitable hermitian operators.
As already mentioned,
the parameters $a$, $b$, $c$, $\alpha$, $\beta$ and $\gamma$ can be assumed
to be small, of order $10^{-19}\ {\rm GeV}$, roughly
of the same order of magnitude of $\epsilon_S \Delta\Gamma$ and
$\epsilon_L \Delta\Gamma$; therefore, one can use an expansion in all these 
small parameters and approximate expressions for the entries of
$\rho(t)$ can be obtained (see [4]).
In particular, one can work out the contributions $\rho_L$ and $\rho_S$  
that correspond to the physical ``long lived'' and ``short lived'' 
neutral kaon states:
$$
\rho_L=N_L\left[\matrix{
\left|\epsilon_L+{2C^*\over\Delta\Gamma_-}\right|^2+{\gamma\over\Delta\Gamma}
-8\left|{C\over\Delta\Gamma_+}\right|^2
-4\,{\cal R}e\left({\epsilon_L C\over\Delta\Gamma}\right)
& \epsilon_L+{2 C^*\over\Delta\Gamma_-}\cr
\phantom{.}&\phantom{.}\cr
\epsilon_L^*+{2 C\over\Delta\Gamma_+} & 1}\right]\ ,
\eqno(5)
$$
and 
$$
\rho_S=N_S\left[\matrix{ 1 & 
\epsilon_S^*+{2 C^*\over\Delta\Gamma_+}\cr
\phantom{.}&\phantom{.}\cr
\epsilon_S+{2 C\over\Delta\Gamma_-} &
\left|\epsilon_S+{2C\over\Delta\Gamma_-}\right|^2 -{\gamma\over\Delta\Gamma}
-8\left|{C\over\Delta\Gamma_+}\right|^2
-4\,{\cal R}e\big({\epsilon_S C^*\over\Delta\Gamma}\big)}\right]\ ,
\eqno(6)
$$
where $C=c+i\beta$ and 
$N_L$, $N_S$ are suitable normalization constants, whose explicit
expressions are irrelevant for the discussion that follows.
Note that the states $\rho_L$ and $\rho_S$ are mixed: the matrices
(5) and (6) are not projectors; only in absence of the contribution
in (4) one recovers the matrices $|K_L\rangle\langle K_L|$ and
$|K_S\rangle\langle K_S|$.

Useful observables are associated with the decays of the neutral
kaons, in particular into pion states.
The amplitudes for the decay of a $K^0$ state into $\pi^+\pi^-$
and $2\pi^0$ final states are usually parametrized as follows,
in terms of the $s$-wave phase-shifts $\delta_i$ and the complex amplitudes
$A_i$, $B_i$, $i=1,\ 2$: [11]
$$
\eqalignno{
&{\cal A}(K^0\rightarrow \pi^+\pi^-)=(A_0+B_0)\, e^{i\delta_0}+{1\over\sqrt2}\,
(A_2+B_2)\, e^{i\delta_2}\ ,&(7a)\cr
&{\cal A}(K^0\rightarrow 2\pi^0)=(A_0+B_0)\, e^{i\delta_0}-\sqrt{2}\,
(A_2+B_2)\, e^{i\delta_2}\ ,&(7b)}
$$
where the indices 0, 2 refers to the total isospin $I$. The amplitudes for
the $\overline{K^0}$ decays are obtained from these with the substitutions:
$A_i\rightarrow A_i^*$ and $B_i\rightarrow -B_i^*$.
The imaginary parts of $A_i$ signals direct $CP$-violation, while a non zero
value for $B_i$ will also break $CPT$ invariance.

To construct the operators that describe these two pion decays in
the formalism of density matrices, one has to pass to the basis
$|K_1\rangle$, $|K_2\rangle$ of definite $CP$.
It is convenient to label the corresponding decay amplitudes
as follows:
$$
\eqalign{
&{\cal A}(K_1\rightarrow \pi^+\pi^-)=X_{+-}\ ,\cr
&{\cal A}(K_1\rightarrow 2\pi^0)=X_{00}\ ,}\qquad\quad
\eqalign{
&{\cal A}(K_2\rightarrow \pi^+\pi^-)=Y_{+-}\ 
{\cal A}(K_1\rightarrow \pi^+\pi^-)\ ,\cr
&{\cal A}(K_2\rightarrow 2\pi^0)=Y_{00}\ 
{\cal A}(K_1\rightarrow 2\pi^0)\ .}
\eqno(8)
$$
The complex parameters $X$ and $Y$ can be easily expressed in terms
of $A_i$, $B_i$ and $\delta_i$ of (7). For the dominant amplitudes, 
one easily finds:
$$
\eqalignno{
&X_{+-}=\sqrt{2}\, \Big[{\cal R}e(A_0)+i{\cal I}m(B_0)\Big]\, e^{i\delta_0}
+\Big[{\cal R}e(A_2)+i{\cal I}m(B_2)\Big]\, e^{i\delta_2}\ ,&(9a)\cr
&X_{00}=\sqrt{2}\, \Big[{\cal R}e(A_0)+i{\cal I}m(B_0)\Big]\, e^{i\delta_0}
-2\Big[{\cal R}e(A_2)+i{\cal I}m(B_2)\Big]\, e^{i\delta_2}\ .&(9b)}
$$
On the other hand, the $K_2$ amplitudes are suppressed, since they involve only
$CP$ and $CPT$ violating terms. For the considerations that follow,
it will be sufficient to work in the approximation that keeps only the
dominant terms in these violating parameters. Then, the amplitude ratios 
$Y_{+-}$ and $Y_{00}$ can be written as:
$$
Y_{+-}=\varepsilon-\epsilon_L+\varepsilon^\prime\ ,\qquad
Y_{00}=\varepsilon-\epsilon_L-2\varepsilon^\prime\ ,
\eqno(10)
$$
where the parameters $\varepsilon$ and $\varepsilon'$ take the familiar
expressions:
$$
\varepsilon=\bigg[{\epsilon_L+\epsilon_S\over2} +
i\, {{\cal I}m(A_0)\over {\cal R}e(A_0)}\bigg]+
\bigg[{\epsilon_L-\epsilon_S\over2} +
{{\cal R}e(B_0)\over {\cal R}e(A_0)}\bigg]\ ,\eqno(11)
$$
and
$$
\varepsilon^\prime= {i e^{i(\delta_2-\delta_0)}\over\sqrt2}\,
{{\cal R}e(A_2)\over {\cal R}e(A_0)}\, \bigg[
{{\cal I}m(A_2)\over {\cal R}e(A_2) }
-{{\cal I}m(A_0)\over {\cal R}e(A_0) }\bigg]
+i\bigg[{{\cal R}e(B_0)\over  {\cal R}e(A_0) }
-{ {\cal R}e(B_2)\over {\cal R}e(A_0)}\bigg]\ .\eqno(12)
$$
The first, second square brackets in (11) and (12) contain the $CP$,
respectively $CPT$, violating parameters arising from the neutral kaon mass and
decay matrices, as well as from their decay amplitudes. The factor
$\omega={\cal R}e(A_2)/{\cal R}e(A_0)$ corresponds to the suppression due to
the $\Delta I=1/2$ rule in the kaon two-pion decays. This ratio is known to be
small; therefore, in presenting the above formulas, first order terms in the
small parameters multiplied by $\omega^2$ have been consistently neglected.
Note that the expressions for $\varepsilon$ and $\varepsilon'$ given in
(11) and (12) are independent from the phase conventions adopted in describing
the kaon states.

Using (8), the operators that describe the 
$\pi^+\pi^-$ and $2\pi^0$ final states and include direct $CP$
and $CPT$ violations are readily found:
$$
{\cal O}_{+-}=|X_{+-}|^2\ \left[\matrix{1&Y_{+-}\cr
                              Y_{+-}^*&|Y_{+-}|^2\cr}\right]\ ,\qquad
{\cal O}_{00}=|X_{00}|^2\ \left[\matrix{1&Y_{00}\cr
                              Y_{00}^*&|Y_{00}|^2\cr}\right]\ .
\eqno(13)
$$
Similar results can be obtained for other decay modes;
for explicit expressions, see [5, 6].

With the help of these matrices, one can now compute various
useful observables of the neutral kaon system.
For example, the decay rates of the the physical states $K_L$ and $K_S$,
represented by $\rho_L$ and $\rho_S$,
into $\pi^+\pi^-$ are simply given by the traces of (13) with the
density matrices (5), (6)
$$
\eqalignno{
&\big|{\cal A}(K_L\rightarrow \pi^+\pi^-)\big|^2\equiv
{\rm Tr}\Big({\cal O}_{+-}\, \rho_L\Big)=
|X_{+-}|^2\, N_L\, R_{+-}^L\ , &(14a)  \cr
&\big|{\cal A}(K_S\rightarrow \pi^+\pi^-)\big|^2\equiv
{\rm Tr}\Big({\cal O}_{+-}\, \rho_S\Big)=  
|X_{+-}|^2\, N_S\, R_{+-}^S\ , &(14b)}
$$
where, up to second order in the small parameters,
$$
\eqalignno{
&R_{+-}^L=
\left|\epsilon_L+{2 C^*\over\Delta\Gamma_-}+Y_{+-}\right|^2
+{\gamma\over\Delta\Gamma}-8\left|{C\over\Delta\Gamma_+}\right|^2
-4\, {\cal R}e\left({\epsilon_L C\over\Delta\Gamma}\right)\ ,&(15a)\cr
&R_{+-}^S=
1+2\,{\cal R}e\left[\left(\epsilon_S+
{2 C\over\Delta\Gamma_-}\right)\,Y_{+-}\right]\ .&(15b)\cr}
$$
Similar expressions hold for the decay into $2\pi^0$; one just needs to
make the substitution $\{+-\}\rightarrow\{00\}$ in the formulas above.

The results in (14) and (15) differ from the standard ones due to the
presence of the non-standard parameters $c$, $\beta$ (via the complex
combination $C$) and $\gamma$. As a consequence, the double ratio:
$$
\left|{{\cal A}(K_L\rightarrow \pi^+\pi^-)\over
{\cal A}(K_S\rightarrow \pi^+\pi^-)}\right|^2 \Bigg/
\left|{{\cal A}(K_L\rightarrow 2\pi^0)\over
{\cal A}(K_S\rightarrow 2\pi^0)}\right|^2\equiv
1+6\,\Sigma\ ,\eqno(16)
$$
is no longer connected in a simple way to the parameter
$\varepsilon'/\varepsilon$. Indeed, to leading order in $CP$ and $CPT$
violation and $\Delta I=1/2$ enhancement, using (10), one finds:
$$
\eqalign{
\Sigma={\cal R}e\left({\varepsilon'\over\varepsilon}\right)\,& 
{ |\varepsilon|^2+
2\,{\cal R}e\big(\varepsilon\, C/\Delta\Gamma_+\big)\over |\varepsilon|^2
+{\cal D} }\cr
&\hskip 2 cm
-2\,{\cal I}m\left({\varepsilon'\over\varepsilon}\right)\,
{ {\cal I}m\big(\varepsilon\, C/\Delta\Gamma_+\big)\over|\varepsilon|^2
+{\cal D} }\ ,\cr}\eqno(17)
$$
where
$$
{\cal D}={\gamma\over\Delta\Gamma}-4\left|{C\over\Delta\Gamma_+}\right|^2
+4\, {\cal R}e\left({\varepsilon\, C\over\Delta\Gamma_+}\right)
-4\, {\cal R}e\left({\epsilon_L C\over\Delta\Gamma}\right)\ .\eqno(18)
$$
Only when $c=\beta=\gamma=\,0$, the expression in (17) reduces to the
standard result: $\Sigma={\cal R}e(\varepsilon'/\varepsilon)$.

This observation is of relevance in view of the fact that
until now experimental informations on $\varepsilon'/\varepsilon$
rely on the measure of the double ratio in (16). 
On general grounds,
one expects the phase of $\varepsilon'$ to be very
close to that of $\varepsilon$,[11] so that
${\cal I}m(\varepsilon'/\varepsilon)$ should be small when compared
with ${\cal R}(\varepsilon'/\varepsilon)$; taking this 
as a working assumption,
one can then neglect the contribution 
from ${\cal I}m(\varepsilon'/\varepsilon)$
in (17). Even in this case, an experimental measure of the
quantity $\Sigma$ will result in a meaningful determination of 
${\cal R}(\varepsilon'/\varepsilon)$ provided estimates on the 
non-standard dissipative parameters $c$, $\beta$ and $\gamma$ are
independently given.

Fortunately, by exploiting the explicit time-dependence of
various neutral kaon observables that are directly accessible to
the experiment, one can get independent evaluations for
all the dissipative parameters appearing in (17), (18). 
For example, the decay rate for a pure $K^0$ initial state
is given by:
$$
\eqalign{
R_{+-}(t)&=
{{\rm Tr}\Big[\rho_{K^0}(t){\cal O}_{+-}\Big]\over
{\rm Tr}\Big[\rho_{K^0}(0){\cal O}_{+-}\Big]}\cr
&=e^{-\gamma_S t}+R_{+-}^L\, e^{-\gamma_L t}+
2\, e^{-\Gamma t}\, |\eta_{+-}|\cos(\Delta m\, t-\phi_{+-})\ ,}\eqno(19)
$$
where $R_{+-}^L$ is the already encountered two-pion decay rate 
for the $K_L$ state, while the combination
$$
\eta_{+-}\equiv|\eta_{+-}|e^{i\phi_{+-}}=
\epsilon_L+Y_{+-}-{2C^*\over\Delta\Gamma_-}=
\varepsilon+\varepsilon'-{2C^*\over\Delta\Gamma_-}\ ,\eqno(20)
$$
determines the interference term. Both $R_{+-}^L$ and $\eta_{+-}$
have been measured using different experimental setups.[12, 13]
Other interesting observables,
are the asymmetries associated with the decay into the final state $f$
of an initial $K^0$ as compared to the corresponding decay 
into $\bar f$ of an initial $\overline{K^0}$. 
All these asymmetries have the general form
$$
A(t)={
{\rm Tr}\Big[\rho_{\bar{K^0}}(t){\cal O}_{\bar f}\Big]-
\Big[\rho_{K^0}(t){\cal O}_f\Big]\over
{\rm Tr}\Big[\rho_{\bar{K^0}}(t){\cal O}_{\bar f}\Big]+
\Big[\rho_{K^0}(t){\cal O}_f\Big]}\ .
\eqno(21)
$$

As discussed in [5], combining the expressions of 
$R_{+-}^L$ and ${\cal R}e(\eta_{+-})$
in $(15a)$ and (20) with asymmetries involving both
three-pion and semileptonic final states one is able 
to obtain bounds on some of the the dissipative parameters
using available experimental data. Since the appearance of [5],
new experimental results have become available, 
in particular from the CPLEAR Collaboration.[14-16] Updating the
discussion in [5] with the new data and the revised world
averages for the parameters $\Delta\Gamma$ and $\Delta m$,[17]
one finally obtains
the following new estimates:
$$
\eqalign{
&c=(0.7 \pm 1.2)\times 10^{-17}\ \hbox{GeV}\ ,\cr
&\beta=(-0.7 \pm 1.3)\times 10^{-17}\ \hbox{GeV}\ ,\cr
&\gamma=(0.1 \pm 22.0)\times 10^{-20}\ \hbox{GeV}\ .}\eqno(22)
$$
Although compatible with zero, these figures can still be used
to give a rough evaluation of the correction to the familiar relation 
$\Sigma={\cal R}e(\varepsilon'/\varepsilon)$,
due to the dissipative parameters.

It should be noticed that the results in (22) have been obtained
neglecting direct $CP$ violating terms,
since they are too small to affect the above figures
for $c$, $\beta$ and $\gamma$; this approximation
is perfectly adequate when computing the dissipative contributions
in (17), since they are already multiplied by a factor $\varepsilon'$.
Furthermore, the validity of the $\Delta S=\Delta Q$ rule has been also
assumed, and direct $CPT$ violating effects in the decay products
have been ignored. With these approximations, $\epsilon_L\simeq\varepsilon$
can be deduced from the measure of the parameter $\eta_{+-}$ in (20).

By writing the relation between the experimentally measured
quantity $\Sigma$ and ${\cal R}e(\varepsilon'/\varepsilon)$ as
$$
{\cal R}e\left({\varepsilon'/\varepsilon}\right)=
\Sigma\, (1-{\cal X})\ ,\eqno(23)
$$
using the above estimates for the the dissipative, 
non-standard parameters, one can obtain a preliminary rough bound
on the correction term
$$
{\cal X}={
4\left|{C\over\Delta\Gamma_+}\right|^2
+4\, {\cal R}e\left({\epsilon_L C\over\Delta\Gamma}\right)
-2\, {\cal R}e\left({\varepsilon\, C\over\Delta\Gamma_+}\right)
-{\gamma\over\Delta\Gamma}
\over
|\varepsilon|^2
+2\, {\cal R}e\left({\varepsilon\, C\over\Delta\Gamma_+}\right)}
\ .\eqno(24)
$$
One explicitly gets:
$$
{\cal X}=0.4\pm 1.8\ .\eqno(25)
$$

The accuracy of the present experimental
results on the neutral kaon system is not high enough to allow a
precise determination of the correction $\cal X$,
which is clearly compatible with zero.
However, the next-generation of dedicated kaon experiments should provide
more complete and precise data, so that the contributions
of dissipative effects to the neutral kaon dynamics and in particular
to the determination of $\varepsilon'/\varepsilon$, if
present, could be directly measured.
If these new experiments will provide
evidence for a nonvanishing ${\cal X}$, the actual value of
$\varepsilon'/\varepsilon$ could be significantly different from
the measured value of the quantity $\Sigma$.

The standard model predicts $\varepsilon'/\varepsilon$ to be non-zero;
however, a precise theoretical determination of its value
is difficult due to cancellations in some hadronic contributions.
The theoretical estimates vary from a few times $10^{-4}$ to about
$10^{-3}$, although with large uncertainties.[18-20]
On the contrary, the KTEV Collaboration has recently
announced a new, very precise
measurement of the double ratio in (16), giving:
$\Sigma=(2.80\pm0.41)\times 10^{-3}$.[21]
Although various mechanisms have been advocated to solve this
apparent discrepancy,[22-24] it is intriguing to observe that the presence
of dissipative effects in the kaon dynamics could account, via
the combination (24), for
the difference between the measured $\Sigma$ and the theoretically
predicted ${\cal R}e(\varepsilon'/\varepsilon)$.

Actually, one can turn the argument around and get an
evaluation of the extra correction $\cal X$
using as inputs the theoretical evaluations of
${\cal R}e(\varepsilon'/\varepsilon)$ and the above experimental
value for $\Sigma$. Although from $\cal X$ alone one can not
directly extract informations on the single parameters
$c$, $\beta$ and $\gamma$, a non-vanishing value of $\cal X$
would clearly signal the presence of dissipative effects
in the neutral kaon dynamics.

The theoretical estimates on ${\cal R}e(\varepsilon'/\varepsilon)$ 
are based on
standard (perturbative and non-perturbative) quantum field techniques,
and ultimately reduce to the evaluation of transition matrix elements
of certain effective operators between kaon and two-pion states.
In principle, the dissipative, non-standard phenomena, whose effects
have been included in the dynamical evolution (2) via the additional
term (4), can also contribute directly to the decay process
of the neutral kaon system; as a consequence, without further analysis
the theoretical prediction of ${\cal R}e(\varepsilon'/\varepsilon)$ should not
be used to obtain a reliable estimate of $\cal X$, since
additional dissipative contributions might have been overlooked.
Fortunately, the effects of dissipative phenomena
to the kaon decay processes have been analyzed, and
found to be negligible for any practical considerations;[25]
in other words, the kaon decay is driven by the weak
interaction and not by dissipative ``gravitational'' 
or ``stringy'' effects. Therefore, at least in principle,
bounds on the existence of the
correction $\cal X$ can be obtained,
by rewriting (23) as:
${\cal X}=1-{\cal R}e(\varepsilon'/\varepsilon)\big/
\Sigma$.

The most accurate theoretical determinations of
${\cal R}e(\varepsilon'/\varepsilon)$ are based on lattice computations,[18]
phenomenological approaches,[19]
and the chiral quark model.[20]
Using the KTEV measure of $\Sigma$, from the results 
for ${\cal R}e(\varepsilon'/\varepsilon)$ given
in Refs.[18, 19, 20], one respectively finds:
$$
{\cal X}=0.79\hbox{$+0.15\atop-0.21$}\ ,\qquad
{\cal X}=\Bigg\{ {0.65\hbox{$+0.16\atop-0.29$}
\atop
0.76\hbox{$+0.13\atop-0.22$}}\ ,\qquad
{\cal X}=0.22\hbox{$+0.68\atop-0.44$}\ .\eqno(26)
$$
The errors in these determinations are based on flat distributions, and give
just a rough idea of the accuracy of the central values.
Although the figures in (26) seem to point towards 
a non-vanishing $\cal X$, one should not regard this result
as an evidence of dissipative effects
in the dynamics of the neutral kaons. Rather, it should be
considered as a rough evaluation of the sensitivity of the
present knowledge of the parameter
$\varepsilon'/\varepsilon$ in testing quantum dynamical
time evolutions of the form given in (2). To improve the bounds
in (26), more accurate inputs are necessary.

Systems of two correlated neutral kaons, 
coming from the decay of a $\phi$-meson,
are also particularly suitable for the study of direct $CP$ violation
phenomena, in view of the possibilities offered by $\phi$-factories.
The $\phi$-meson has spin one, and therefore
its decay into two spinless bosons produces
an antisymmetric spatial state. In the $\phi$ rest frame, the two neutral kaons
are produced flying apart with opposite momenta; in the basis $|K_1\rangle$,
$|K_2\rangle$, the resulting state can be described by:
$$
|\psi_A\rangle= {1\over\sqrt2}\Big(|K_1,-p\rangle \otimes  |K_2,p\rangle -
|K_2,-p\rangle \otimes  |K_1,p\rangle\Big)\ .
\eqno(27)
$$
The corresponding density operator $\rho_A$ is a now $4\times 4$ matrix,
and its time-evolution can be obtained by assuming that, once
produced in a $\phi$ decay, the kaons evolve in time each according to the
completely positive map generated by the equation (2).[6]

The typical observables that can be studied in such physical situations
are double decay rates, {\it i.e.} the probabilities that a kaon decays
into a final state $f_1$ at proper time $\tau_1$, while the other kaon 
decays into the final state $f_2$ at proper time $\tau_2$:
$$
{\cal G}(f_1,\tau_1; f_2,\tau_2)\equiv 
\hbox{Tr}\Big[\Big({\cal O}_{f_1}\otimes{\cal O}_{f_2}\Big) 
\rho_A(\tau_1,\tau_2)\Big]\ ;
\eqno(28)
$$
as before, the operators ${\cal O}_{f_1}$ and ${\cal O}_{f_2}$
are the $2\times 2$ hermitian matrices describing the final decay states.
However, much of the analysis at $\phi$-factories is carried out using
integrated distributions at fixed time interval $\tau=\tau_1-\tau_2$.
One then deals with single-time distributions, defined by:
$$
{\mit\Gamma}(f_1,f_2;\tau)=\int_0^\infty dt\, {\cal G}(f_1,t+\tau;f_2,t)\ ,
\qquad \tau>0\ .\eqno(29)
$$
Starting with these integrated probabilities, one can form asymmetries
that are sensitive to various parameters in the theory. In particular,
in the case of two-pion final states, the following observable
is particularly useful for the determination of 
$\varepsilon'/\varepsilon$:[26]
$$
{\cal A}_{\varepsilon^\prime}(\tau)={
{\mit\Gamma}(\pi^+\pi^-,2\pi^0;\tau) - 
{\mit\Gamma}(2\pi^0,\pi^+\pi^-;\tau)\over
{\mit\Gamma}(\pi^+\pi^-,2\pi^0;\tau) + 
{\mit\Gamma}(2\pi^0,\pi^+\pi^-;\tau) }\ .
\eqno(30)
$$
Indeed, one can show that, to first order in all small parameters:
$$
{\cal A}_{\varepsilon^\prime}(\tau)=3\, 
{\cal R}e\Big({\varepsilon^\prime\over\varepsilon}\Big)\ 
{N_1(\tau)\over D(\tau)}
-3\, {\cal I}m\Big({\varepsilon^\prime\over\varepsilon}\Big)\ 
{N_2(\tau)\over D(\tau)}\ .\eqno(31)
$$
The $\tau$-dependent coefficients $N_1$, $N_2$ and $D$ are 
functions of $\varepsilon$ and of the
dissipative parameters $c$, $\beta$ and $\gamma$;
their explicit expressions are collected in the Appendix.
The clear advantage of using the asymmetry 
${\cal A}_{\varepsilon^\prime}$ to determine
the value of $\varepsilon'/\varepsilon$ in comparison to the
double ratio (16) is that, in principle, both real and imaginary
part can be extracted from the time behaviour of (31).
Due to the presence of the dissipative parameters however,
this appears to be much more problematic than in the
standard case; once more, a meaningful determination of
$\varepsilon'/\varepsilon$ is possible provided independent estimates
on $c$, $\beta$ and $\gamma$ are obtained from
the measure of other independent asymmetries.
This is particularly evident if one looks at the large-time
limit $(\tau\gg1/\gamma_S)$ of (31); indeed, in this limit,
the expression of ${\cal A}_{\varepsilon'}/3$ coincides 
with that of $\Sigma$ given in (17), and the considerations presented 
there in discussing the case of a single kaon system apply also here.

In conclusion, dissipative effects in the dynamics of both single
and correlated neutral kaon systems could affect the precise determination
of the ratio $\varepsilon'/\varepsilon$; future dedicated experiments
on kaon physics will certainly provide stringent bounds on these
dissipative effects, thus clarifying their role in the analysis
of the direct $CP$ violating $K^0$-$\overline{K^0}$ pion-decays.

\vfill\eject

{\bf Appendix}
\medskip

For completeness, we collect here the explicit form of the three
coefficients $N_1$, $N_2$ and $D$ that appear in the
expression (31) for the asymmetry ${\cal A}_{\varepsilon'}$;
they have been discussed also in [6], although using
different approximations.

$$
\eqalignno{
&N_1(\tau)=e^{-\gamma_L\tau}\bigg[ |\varepsilon|^2
+2\, {\cal R}e\bigg({\varepsilon\,C\over\Delta\Gamma_+}\bigg)\bigg]
-e^{-\gamma_S\tau}\bigg[|\varepsilon|^2
-2\,{\cal R}e\bigg({\varepsilon\, C\over\Delta\Gamma_+}\,
{\gamma_L+\Gamma_-\over\gamma_S+\Gamma_+}\bigg)\bigg]\cr
&\hskip 2cm -4e^{-\Gamma\tau}\bigg[{\cal R}e\bigg(e^{-i\Delta m\tau}
{\varepsilon\,C\over\Delta\Gamma_+}\,
{2\Gamma\over\gamma_S+\Gamma_+}\bigg)
-{\cal I}m\bigg({\varepsilon\,C\over\Delta\Gamma_+}\bigg)
\,\sin(\Delta m\tau)\bigg] &(A.1)\cr
\cr\cr
&N_2(\tau)=2 e^{-\gamma_L\tau}\  
{\cal I}m\bigg({\varepsilon\,C\over\Delta\Gamma_+}\bigg)
+2 e^{-\gamma_S\tau}\
{\cal I}m\bigg({\varepsilon\, C\over\Delta\Gamma_+}\,
{\gamma_L+\Gamma_-\over\gamma_S+\Gamma_+}\bigg)\cr
&\hskip .5cm -2e^{-\Gamma\tau}\bigg\{{\cal I}m\bigg[e^{-i\Delta m\tau}
\bigg(|\varepsilon|^2+
{2\varepsilon\,C\over\Delta\Gamma_+}\,
{2\Gamma\over\gamma_S+\Gamma_+}\bigg)\bigg]+
2\,{\cal R}e\bigg({\varepsilon\,C\over\Delta\Gamma_+}\bigg)
\,\sin(\Delta m\tau)\bigg\} \cr
&&(A.2)\cr
\cr\cr
&D(\tau)=e^{-\gamma_L\tau}\bigg[|\varepsilon|^2+
{\gamma\over\Delta\Gamma}-4\left|{C\over\Delta\Gamma_+}\right|^2-
4\, {\cal R}e\bigg({\epsilon_L C\over\Delta\Gamma}\bigg)
+4\, {\cal R}e\bigg({\varepsilon\,C\over\Delta\Gamma_+}\bigg)\bigg]\cr
&\hskip 1cm\ +e^{-\gamma_S\tau}\bigg\{|\varepsilon|^2-
{\gamma_L\over\gamma_S}\bigg[{\gamma\over\Delta\Gamma}
-4\,{\cal R}e\bigg({\epsilon_L C\over\Delta\Gamma}\bigg)-
8\left|{C\over\Delta\Gamma_+}\right|^2\bigg]\cr
&\hskip 2cm -4\, {\cal R}e\bigg({\varepsilon\, C\over\Delta\Gamma_+}\,
{\gamma_L+\Gamma_-\over\gamma_S+\Gamma_+}\bigg)
+4\bigg|{C\over\Delta\Gamma_+}\bigg|^2
\bigg[{8\,\Gamma(\Gamma+\gamma_S)\over(\Gamma+\gamma_S)^2 +(\Delta m)^2}
-3\bigg]\bigg\}\cr
&\hskip 1cm -2e^{-\Gamma\tau}\bigg\{ {\cal R}e\bigg[e^{-i\Delta m\tau}\bigg(
|\varepsilon|^2+{4\,\varepsilon\,C\over\Delta\Gamma_+}\,
{2\,\Gamma\over\gamma_S+\Gamma_+}
-4\bigg|{C\over\Delta\Gamma_+}\bigg|^2
{\gamma_L+\Gamma_-\over\gamma_S+\Gamma_+}\bigg)\bigg]\cr
&\hskip 7cm-4\,{\cal R}e\bigg({\varepsilon\,C\over\Delta\Gamma_+}\bigg)\,
\cos(\Delta m\tau)\bigg\} \cr
&&(A.3)}
$$

\vfill\eject


\centerline{\bf References}
\bigskip

\item{1.} R. Alicki and K. Lendi, {\it Quantum Dynamical Semigroups and 
Applications}, Lect. Notes Phys. {\bf 286}, (Springer-Verlag, Berlin, 1987)
\smallskip
\item{2.} V. Gorini, A. Frigerio, M. Verri, A. Kossakowski and
E.C.G. Surdarshan, Rep. Math. Phys. {\bf 13} (1978) 149 
\smallskip
\item{3.} H. Spohn, Rev. Mod. Phys. {\bf 53} (1980) 569
\smallskip
\item{4.} F. Benatti and R. Floreanini, Nucl. Phys. {\bf B488} (1997) 335
\smallskip
\item{5.} F. Benatti and R. Floreanini, Phys. Lett. {\bf B401} (1997) 337
\smallskip
\item{6.} F. Benatti and R. Floreanini, Nucl. Phys. {\bf B511} (1998) 550
\smallskip
\item{7.} S. Hawking, Comm. Math. Phys. {\bf 87} (1983) 395; Phys. Rev. D
{\bf 37} (1988) 904; Phys. Rev. D {\bf 53} (1996) 3099;
S. Hawking and C. Hunter, Phys. Rev. D {\bf 59} (1999) 044025
\smallskip
\item{8.} F. Benatti and R. Floreanini, Ann. of Phys. {\bf 273} (1999) 58,
{\tt hep-th/9811196}
\smallskip
\item{9.} F. Benatti and R. Floreanini, 
Mod. Phys. Lett. {\bf A12} (1997) 1465; 
Banach Center Publications, {\bf 43} (1998) 71; 
Comment on ``Searching for evolutions 
of pure states into mixed states in the two-state system $K$-$\overline{K}$'',
{\tt hep-ph/9806450}
\smallskip
\item{10.} T.D. Lee and C.S. Wu, Ann. Rev. Nucl. Sci. {\bf 16} (1966) 511
\smallskip
\item{11.} L. Maiani, $CP$ and $CPT$ violation in neutral kaon decay,
in {\it The Second Da$\,\mit\Phi$ne Physics Handbook}, 
L. Maiani, G. Pancheri and N. Paver, eds., (INFN, Frascati, 1995)
\smallskip
\item{12.} C. Geweniger {\it et al.}, Phys. Lett. {\bf B48} (1974) 487
\smallskip
\item{13.} The CPLEAR Collaboration, Phys. Lett. {\bf B369} (1996) 367
\smallskip
\item{14.} The CPLEAR Collaboration, Phys. Lett. {\bf B444} (1998) 43
\smallskip
\item{15.} The CPLEAR Collaboration, Eur. Phys. J. C{\bf 5} (1998) 389
\smallskip
\item{16.} P. Kokkas, talk presented at the {\it XXIX International Conference
on High Energy Physics}, Vancouver, July, 1998
\smallskip
\item{17.} Particle Data Group, Eur. Phys. J. C{\bf 3} (1998) 1
\smallskip
\item{18.} M. Ciuchini, Nucl. Phys. Proc. Suppl. {\bf 59} (1997) 149,
and references therein
\smallskip
\item{19.} S. Bosch, A.J. Buras, M. Gorbahn, S. J\"ager,
M.Jamin, M.E. Lautenbacher and\hfill\break 
L. Silvestrini, Standard model confronting new results for 
$\varepsilon'/\varepsilon$, {\tt hep-th/9904408}
\smallskip
\item{20.} S. Bertolini, M.Fabbrichesi and J.O. Eeg, Estimating
$\varepsilon'/\varepsilon$, {\tt hep-th/9802405}
\smallskip
\item{21.} The KTEV Collaboration,\hfill\break
{\tt http://fnphyx-www.fnal.gov/experiments/ktev/epsprime/epsprime.html}
\smallskip
\item{22.} Y.-Y. Keum, U. Nierste and A.I. Sanda, A short look at
$\varepsilon'/\varepsilon$, {\tt hep-th/9903230}
\smallskip
\item{23.} X.-G. He, Contribution to $\varepsilon'/\varepsilon$
from anomalous gauge couplings, {\tt hep-th/9903242}
\smallskip
\item{24.} A. Masiero and H. Murayama, Can $\varepsilon'/\varepsilon$ be
supersymmetric?, {\tt hep-th/9903363}
\smallskip
\item{25.} F. Benatti and R. Floreanini, Phys. Lett. {\bf B428} (1998) 149
\smallskip
\item{26.} G. D'Ambrosio, G. Isidori and A. Pugliese, $CP$ and $CPT$
measurements at Da$\Phi$ne, in 
{\it The Second Da$\,\mit\Phi$ne Physics Handbook}, 
L. Maiani, G. Pancheri and N. Paver, eds., (INFN, Frascati, 1995)

\bye